\documentclass[fleqn,usenatbib]{mnras}

\usepackage[T1]{fontenc}
\usepackage{ae,aecompl}

\usepackage{graphicx}	% Including figure files
\usepackage{amsmath}	% Advanced maths commands
\usepackage{amssymb}	% Extra maths symbols
\newcommand{\epse}{\epsilon_{e}}
\newcommand{\epsB}{\epsilon_{B}}

\newcommand{\thetaobs}{\theta_v}
\newcommand{\Mej}{M_{\mathrm{ej}}}
\newcommand{\umin}{u_{\mathrm{min}}}
\newcommand{\umax}{u_{\mathrm{max}}}
\newcommand{\Einj}{E_{\mathrm{inj}}}

\title[A year of GW 170817]{
A year in the life of GW170817: 
the rise and fall of a structured jet from a binary neutron star merger
}

\author[Troja et al.]{
E. Troja$^{1,2}$\thanks{E-mail: eleonora@umd.edu}, H. van Eerten$^{3}$,
G. Ryan$^{1}$,
R. Ricci$^{4}$,
J.~M.~Burgess$^{5,6}$,
M. H.~Wieringa$^{7}$,
\newauthor
L. Piro$^{8}$,
S. B. Cenko$^{2,1}$,
T. Sakamoto$^{9}$ 
\\
$^{1}$ Department of Astronomy, University of Maryland, College Park, MD 20742-4111, USA \\
$^{2}$Astrophysics Science 
Division, NASA Goddard Space Flight Center, 8800 Greenbelt Rd, Greenbelt, MD 20771, USA\\
$^{3}$Department of Physics, University of Bath, Claverton Down, Bath BA2 7AY, United Kingdom\\
$^{4}$INAF-Istituto di Radioastronomia, Via Gobetti 101, I-40129, Italy\\
 $^{5}$Max-Planck-Institut fur extraterrestrische Physik, Giessenbachstrasse 1, D-85748 Garching, Germany\\
$^{6}$ Excellence Cluster Universe, Technische Universit{\"a}t M{\"u}nchen, Boltzmannstra{\ss}e 2, 85748 Garching,Germany\\
$^{7}$CSIRO Astronomy and Space Science, P.O. Box 76, Epping NSW 1710, Australia\\
$^{8}$INAF, Istituto di Astrofisica e Planetologia Spaziali, via Fosso del Cavaliere 100, 00133 Rome, Italy\\
$^{9}$%Department of Physics and Mathematics, 
Aoyama Gakuin University, 5-10-1 Fuchinobe, Chuoku, Sagamiharashi Kanagawa 252-5258, Japan \\
}
\date{Accepted 12 August 2019. Received 5 August 2019; in original form 21 August 2018}

\pubyear{2019}

\begin{document}

\pagerange{\pageref{firstpage}--\pageref{lastpage}}
\maketitle
\label{firstpage}

% Abstract of the paper
\begin{abstract}
We present the results of our year-long afterglow monitoring of GW170817, the first binary neutron star (NS) merger
detected by advanced LIGO and advanced Virgo. 
New observations with the Australian Telescope Compact Array (ATCA) and the {\it Chandra X-ray Telescope} were used to constrain its late-time behavior. 
The broadband emission, from radio to X-rays, is well-described by a simple power-law spectrum with index $\beta$\,$\sim$0.585 at all epochs. 
After an initial shallow rise $\propto$\,$t^{0.9}$, 
the afterglow displayed a smooth turn-over, reaching a peak X-ray luminosity of $L_X$$\approx$5$\times$$10^{39}$\,erg\,s$^{-1}$ at 160 d,  and has now entered a phase of rapid decline, 
approximately $\propto$\,$t^{-2}$. 
The latest temporal trend challenges most models of choked jet/cocoon systems, and is instead consistent with the emergence of a relativistic structured jet seen at an angle of $\approx$22$^{\circ}$ from its axis.
Within such model, the properties of the explosion (such as its blastwave energy 
$E_K\approx2\times10^{50}$\,erg, jet width $\theta_c$\,$\approx$4$^{\circ}$, and ambient density $n$\,$\approx$3$\times$\,$10^{-3}$\,cm$^{-3}$)
 fit well within the range of properties of cosmological short GRBs.

\end{abstract}

% Select between one and six entries from the list of approved keywords.
% Don't make up new ones.

%%%%%%%%%%%%%%%%%%%%%%%%%%%%%%%%%%%%%%%%%%%%%%%%%%

%%%%%%%%%%%%%%%%% BODY OF PAPER %%%%%%%%%%%%%%%%%%

\section{Introduction}

On August 17$^{\rm th}$, 2017 the Advanced LIGO interferometers detected the first 
gravitational wave (GW) signal from a binary neutron star (NS) merger, GW170817, 
followed 1.7 s later by a short duration gamma-ray burst, GRB170817A \citep{LVCGBM}.
Located in the elliptical galaxy NGC4993 at a distance of $\sim$40 Mpc, 
 GRB170817A was an atypical sub-luminous explosion.
An X-ray afterglow was detected 9 days after the merger \citep{Troja2017}.
A second set of observations, performed $\sim$15 days post-merger, revealed that the emission 
was not fading, as standard GRB afterglows, but was instead rising at a slow rate \citep{Troja2017,Haggard2017}.
The radio afterglow, detected at 16 days \citep{Hallinan2017},
continued to rise in brightness \citep{Mooley2018}, as later confirmed by
X-ray and optical observations \citep{Troja2018,DAvanzo2018,Lyman2018,Margutti2018}. 

The delayed afterglow onset and  low-luminosity of the $\gamma$-ray { signal}
could be explained if the jet was observed at an angle (off-axis) of $\approx$15$^{\circ}$-30$^{\circ}$.
Whereas a standard uniform jet viewed off-axis could account 
for the early afterglow emission, \citet{Troja2017} and \citet{Kasliwal2017} noted that it could not account for the observed gamma-ray signal and proposed two alternative models:  a structured jet, i.e. a jet with an {\it angular} profile
of Lorentz factors and energy \citep[see also][]{LVCGBM,kathi18}, and a mildly-relativistic isotropic cocoon \citep[see also][]{Lazzati2017b,Kasliwal2017}. 
In the latter model, the jet may never emerge from the merger ejecta (choked jet). 

The subsequent rebrightening ruled out both the uniform jet 
and the simple cocoon models, which predict a sharp afterglow rise. 
It was instead consistent with an off-axis structured jet \citep{Troja2017} and 
a cocoon with energy injection \citep{Mooley2018}, 
characterized by a {\it radial} profile of ejecta velocities. 
In \cite{Troja2018} we developed semi-analytical models for both the structured jet and the quasi-spherical cocoon with energy injection, and showed that they describe the broadband afterglow evolution during the first six months (from the afterglow onset to its peak) equally well.
This is confirmed by numerical simulations of relativistic jets \citep{Lazzati2018,Xie2018} and choked jets \citep{Nakar2018}.

Several tests were discussed to distinguish between these two competing models
\citep[e.g.][]{GillGranot2018,Nakar2018}. 
\citet{Corsi2018} used the afterglow polarization to probe the outflow geometry (collimated vs. nearly isotropic), but the results were not constraining. 
 \citet{Ghirlanda2018} and \citet{Mooley2018superluminal} used Very Long Baseline Interferometry (VLBI) to image the radio counterpart, and
concluded that the compact source size ($\lesssim$2 mas) and its apparent superluminal motion favor the emergence of a relativistic jet core. 
A third and independent way to probe the outflow structure is to follow its late-time temporal evolution. In the case of a cocoon-dominated emission, the afterglow had been predicted to follow a shallow decay ($t^{-\alpha}$) with $\alpha$\,$\sim$1.0-1.2 \citep{Troja2018} for a quasi-spherical outflow, and $\alpha$\,$\sim$1.35 for a wide-angled cocoon \citep{LambMandelResmi2018}.
A relativistic jet is instead expected to resemble a standard on-axis explosion at late-times, thus displaying a post-jet-break decay of $\alpha \sim 2.5$ \citep{vanEertenMacFadyen2013}.

Here, we
present the results of our year-long observing campaign of GW170817, carried out with the Australian Telescope Compact Array (ATCA) in the radio, {\it Hubble Space Telescope} (HST) in the optical, 
the {\it Chandra} X-ray telescope and {\it XMM-Newton} in the X-rays. 
Our latest observations show no signs of spectral evolution (Sect. \ref{spectral_properties_section}) and a rapid decline of the afterglow emission (Sect. \ref{temporal_properties_section}), 
systematically faster than 
cocoon-dominated/choked jet models from the literature (Sect. \ref{decline_results_section}).
The rich broadband dataset allows us to tightly constrain the afterglow parameters,
and to compare the explosion properties of GW170817 to canonical short GRBs (Sect. \ref{comparison_section}).

\section{Observations and data analysis}

Our earlier observations were presented in \cite{Troja2017, Troja2018} and \cite{Piro2018}. 
To these, we add a new series of observations tracking the post-peak afterglow evolution.
Table~\ref{tab:obs} lists the latest unpublished data set, including our radio monitoring with ATCA (PI: Piro, Murphy) and X-ray observations with {\it Chandra}, carried out under our approved General Observer program (20500691; PI: Troja). Data were reduced and analyzed as detailed in \cite{Troja2018,Piro2018}. {In the latest {\it Chandra} observation, the source is detected at a count-rate of (4.9$\pm$0.9)$\times$10$^{-4}$\,cts\,s$^{-1}$ in the 0.3-8.0~keV band, corresponding to an unabsorbed flux of (6.70$\pm$0.13)$\times$10$^{-15}$\,erg\,cm$^{-2}$\,s$^{-1}$. }
We adopted a standard $\Lambda$CDM cosmology \citep{Planck2018}. Unless otherwise stated, the quoted errors are at the 68\% confidence level, and upper limits are at the 3~$\sigma$ confidence level. 

\begin{figure}

\includegraphics[width=1.05\columnwidth]{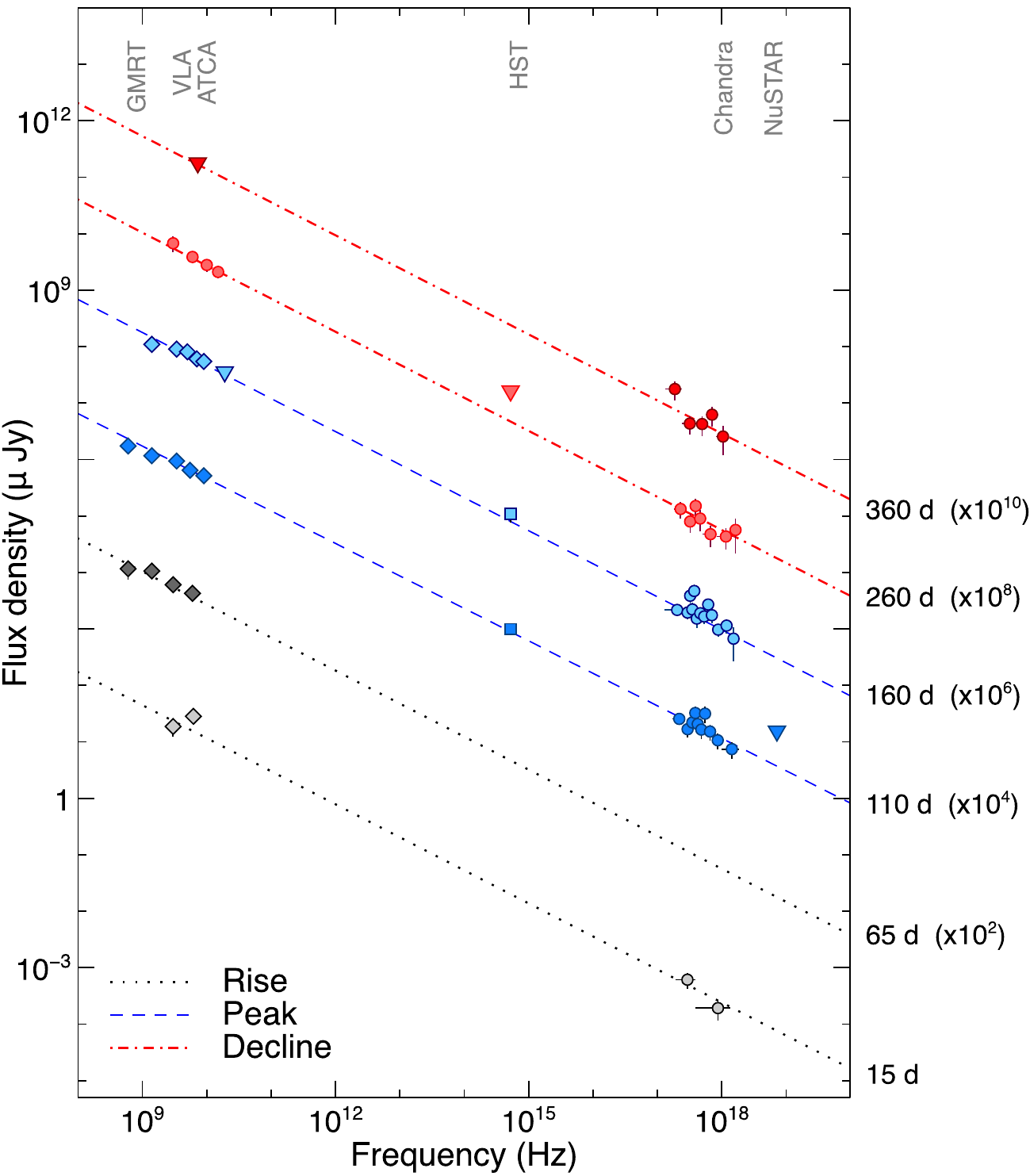}
     \caption{Temporal evolution of the afterglow spectral energy distribution. 
     A single power-law segment can describe the broadband spectrum 
     during the different afterglow phases (rise, peak and decline). 
     At all times, a simple power-law model provides a good fit of the data.}
     \label{fig:sedx}
     \vspace{-0.5cm}
\end{figure}

\subsection{Spectral properties} \label{spectral_properties_section}

The latest epoch of X-ray observations shows a simple power-law spectrum with index $\beta$=0.8$\pm$0.4, consistent with previous measurements, and 
with the spectral index 0.4$\pm$0.3 from the late time ($t>$220 d) radio data. 
Figure~\ref{fig:sedx} shows that, at all epochs, the broadband spectrum
can be fit with a simple power-law model with spectral index $\beta$=0.585$\pm$0.005 and no intrinsic absorption in addition to the Galactic value $N_H$=7.6$\times$10$^{20}$\,cm$^{-2}$.
The lack of any significant spectral variation on such long timescales is remarkable. 
In GRB afterglows, a steepening of the X-ray spectrum due to the gradual decrease of the cooling frequency $\nu_c$ is commonly detected within a few days. 
Since the cooling break is a smooth spectral feature, we used a curved afterglow spectrum \citep{GranotSari2002} to fit the data, and constrain its location\footnote{ The spectrum was fit with a curved spectrum leaving the cooling frequency as a free parameter. The best fit statistics C$_0$ was recorded, then the location of $\nu_c$ was changed until the variation in the fit statistics was equal to $C_0$+2.706.}. 
We derived $\nu_c \gtrsim$1~keV (90\% confidence level) at 260 d after the merger and $\nu_c \gtrsim$0.1~keV (90\% confidence level) at 360 d.

\begin{figure}
\includegraphics[width=0.95\columnwidth]{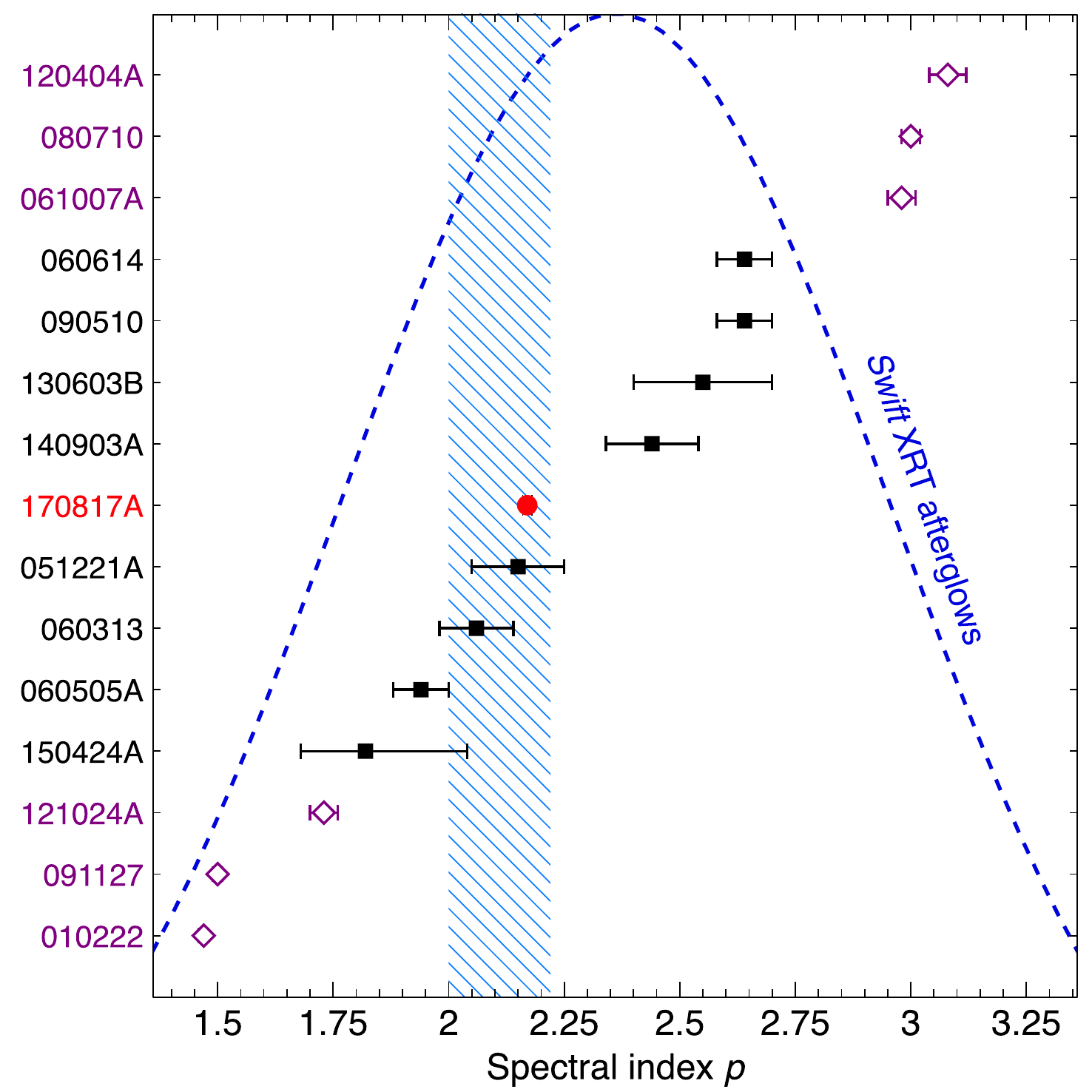}
     \caption{Spectral index $p$ of the shock-accelerated electrons for GW170817/GRB170817A (filled circle) and a sample of
     of short GRBs (filled squares) and long GRBs (open diamonds) with good 
     afterglow constraints. The dashed line shows the distribution inferred from XRT afterglows \citep{Curran2010}. 
     The hatched area shows the range of values predicted by shock theory. 
     Data are from: \citet{Roming2005,Soderberg2015,Mundell2007,Resmi2008,Xu2009,Kruler2009,Kumar2010,Troja2012,Troja2016, Fong2015,Varela2016,Knust2017}.
     }
     \label{fig:p}
     \vspace{-0.3cm}
 \end{figure}

For a synchrotron spectrum with $\nu_m < \nu_{obs} < \nu_c$, the measured spectral index is related to the spectral index $p$ of the emitting electrons \citep[e.g.][]{GranotSari2002} as $p$ = 2$\beta$ +1 = 2.170 $\pm$ 0.010. 
Figure \ref{fig:p} compares this value with a sample of well-constrained short and long GRB spectra. It shows that, among short GRB spectra, GW170817/GRB170817A  represents the most precise measurement obtained so far. Such  precision is rare, but not unprecedented among long GRB afterglows.

It is tempting to interpret the accurately determined value of $p$ as an intermediate between relativistic and non-relativistic shock acceleration, based on theoretical considerations of plausible mechanisms \citep{Kirk2000, Achterberg2001, Spitkovsky2008}, implying $\Gamma$\,$\approx$2-10 for the emitting material \citep{Margutti2018}.
On the other hand, various well constrained short GRB $p$-values lie outside this theoretical range (Fig. \ref{fig:p}), the $p$-value distribution for the larger sample of (long) GRBs does not appear consistent with a universal value for $p$ \citep{ShenKumarRobinson2006, Curran2010}, nor is there generally any evidence for evolution of $p$ from multi-epoch spectral energy distributions (SEDs) (Fig. \ref{fig:sedx}, also e.g. \citealt{Varela2016}) or light curve slopes. In view of these features of the general sample of long and short GRBs (as well as other synchrotron sources, such as blazars), direct interpretation of $p \approx 2.17$ in terms of shock-acceleration theory might be premature.

\vspace{-0.5cm}
\subsection{Temporal properties}
\label{temporal_properties_section}

We simultaneously fit the multi-color { (X-ray, optical, radio)} light curves by adopting a simple power-law function
in the spectral domain, and a smoothly broken power-law in the temporal domain \citep{beuermann99}.
The functional form is: 
$F(\nu , t) \propto \nu^{- \beta} 
\left[(t/t_p)^{-\kappa\alpha_1} +(t/t_p)^{\kappa\alpha_2} \right]^{-1/\kappa}$, 
where $\beta$ is the spectral index, $\alpha_1$ and $\alpha_2$ are the 
rise and decay slopes, $t_p$ is the peak time, and $\kappa$ is the smoothness parameter. 

We did not impose an achromatic behavior. Instead, the temporal properties were modeled as a hierarchy where the parameters for each wavelength have an independent value, sampled from a hyper distribution. 
Variations in the hierarchy would be integrated out of the posterior if justified by the data.

\begin{figure}
    \centering
    \includegraphics[width=0.99\columnwidth]{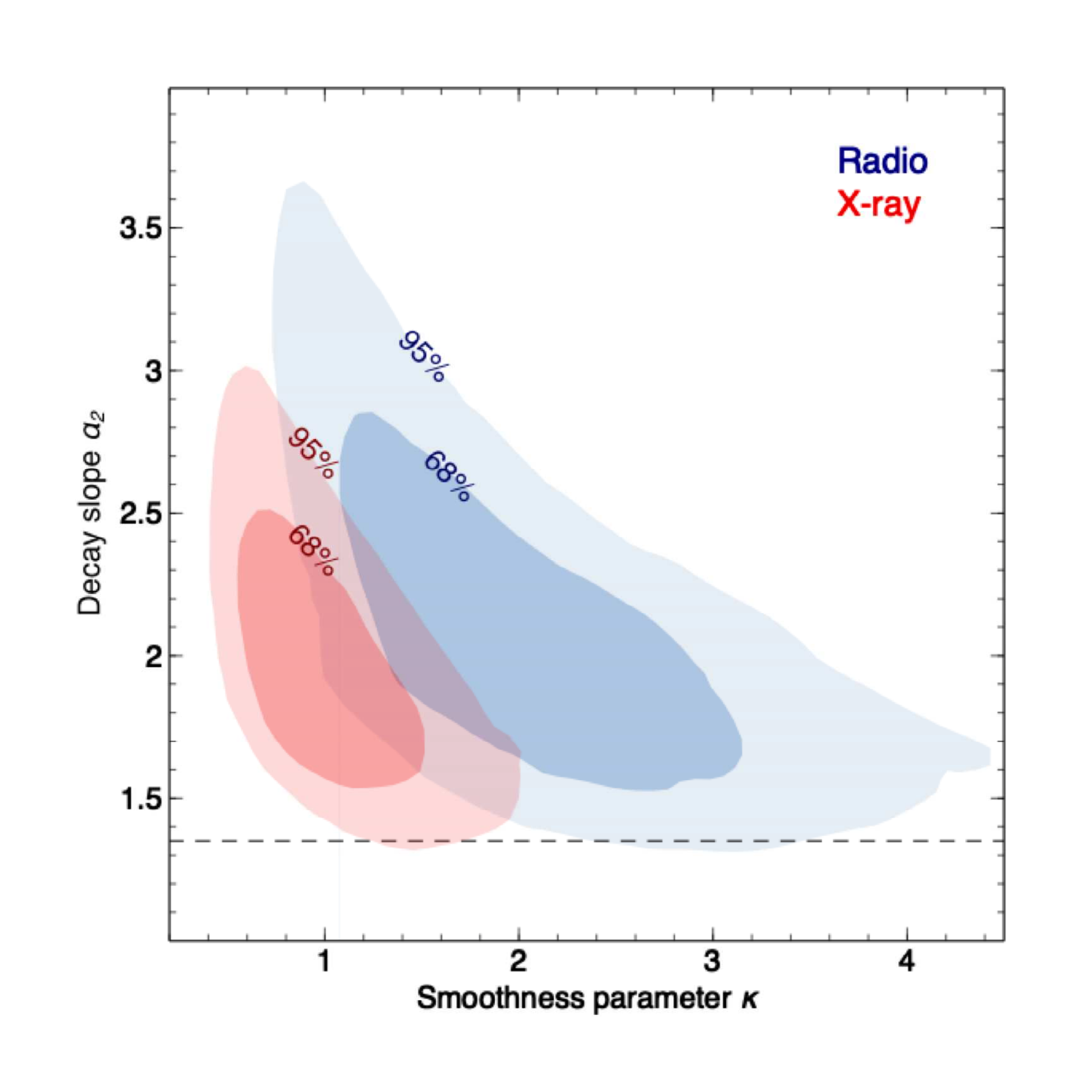}
    \caption{Joint marginal distribution of the decay index ($\alpha_2$) and smoothness parameter ($s$) 
    for the radio (blue) and X-ray (red) light curves. 
    The dashed line corresponds to $\alpha_2=1.35$, the steepest value predicted by choked jet/cocoon models.}
    \label{fig:splot}
\end{figure}

Scale-family distributions were given log-normal priors
and all other parameters were given normal priors. To fit the  model, we employ the NUTS variant of Hamiltonian Monte Carlo via the Stan modeling language \citep{Carpenter2017}.
The peak time and rise slope are well constrained to $t_p$=164$\pm$12 d and $\alpha_1$=0.90$\pm$0.06. 
Whereas the optical afterglow is poorly sampled, the X-ray and radio light curves allow for better constraints on the decay slope, { $\alpha_2$=2.0$^{+0.8}_{-0.5}$} (Figure~\ref{fig:splot}), and are both consistent with a rapid decline of the afterglow flux.  
The best fit model and full corner plot is reported in the Supplementary Material (Figure~\ref{fig:sup1}). 

\begin{figure*}
\includegraphics[width=1.95\columnwidth]{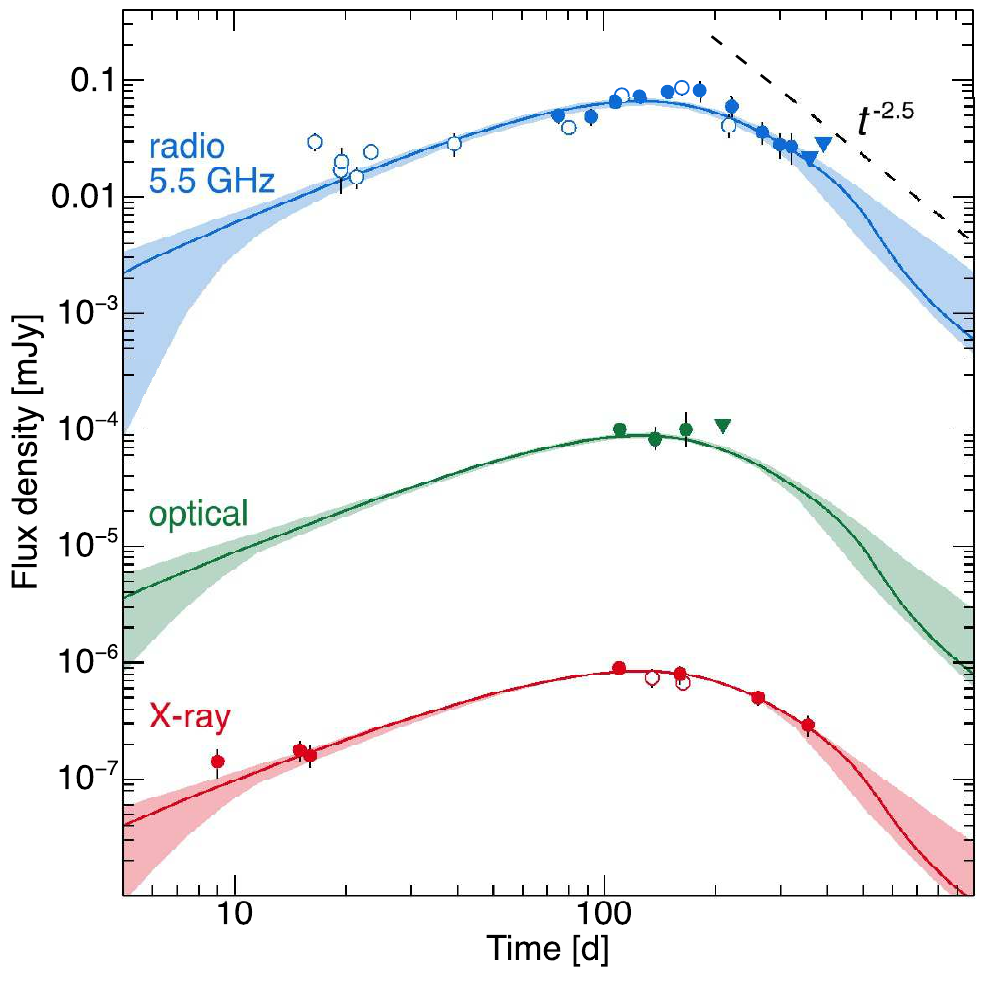}
     \caption{Multi-wavelength afterglow light curves overlaid with the Gaussian jet best fit model (solid line) and its 68\% uncertainty range (shaded areas). 
     Radio data are from ATCA  (filled symbols) and VLA (open symbols) observations. 
     X-ray data are from Chandra (filled symbols) and XMM-Newton (open symbols) observations. 
     Downward triangles are 3 $\sigma$ upper limits. The dashed line shows the expected asymptotic decline $\propto t^{-2.5}$.
     Data were collected from: \citealt{Troja2017}, \citealt{Troja2018}, \citealt{Piro2018}, \citealt{Hallinan2017}, \citealt{Lyman2018}, \citealt{Resmi2018}, \citealt{Margutti2018}, \citealt{Mooley2018}, and \citealt{Alexander2018}.}
     \label{fig:lc}
     \vspace{-0.3cm}
\end{figure*}

\vspace{-0.5cm}
\subsection{Modeling}
\label{modeling_section}

We directly fit two semi-analytical models for structured outflows to the data, following the description in \cite{Troja2018}. 
The off-axis structured jet model assumes a Gaussian energy profile $E \propto \exp \left(-\theta^2/2\theta_c^2\right)$ up to a truncating angle $\theta_w$.  
The jet is fully determined by a set of eight parameters $\Theta_{\mathrm{jet}} = \{ E_0, n, \epse, \epsB\, p, \theta_c, \theta_w, \thetaobs\}$, where $E_0$ is the on-axis isotropic-equivalent kinetic energy of the blast wave, $n$ the circumburst density,  $\epse$ the electron energy fraction, $\epsB$ the magnetic energy fraction and $\thetaobs$ the angle between the jet-axis and the observer's line of sight. 
Following \cite{Mooley2018}, we also fit a cocoon model with a velocity stratification of the ejecta to allow for a slower rise and late turnover. The total amount of energy in the slower ejecta above a particular four-velocity $u$ is modelled as a power-law $E(>u) = \Einj u^{-k}$.
This model requires nine parameters $\Theta_{\mathrm{cocoon}} = \{ \Einj, n, p, \epse, \epsB\, \Mej, \umax, \umin,  k \}$, where $\umax$ is the maximum ejecta four-velocity, $\umin$ the minimum ejecta four-velocity, and $\Mej$ the initial cocoon ejecta mass with speed $\umax$. 

As described in \cite{Troja2018}, our Bayesian fit procedure utilizes the \emph{emcee} Markov-chain Monte Carlo package \citep{Foreman-Mackey2013}. For the structured jet we also include the GW constraints on the orientation $\iota$ of the system \citep{gwhubble} in our prior for $\thetaobs$. The results of the MCMC analysis are summarized in Table~\ref{tab:MCMC}. The best fit jet models is shown in Figure~\ref{fig:lc}. For the corner plot see the Supplementary data (Figure~\ref{fig:corner}).

\begin{figure*}
\includegraphics[width=1.99\columnwidth]{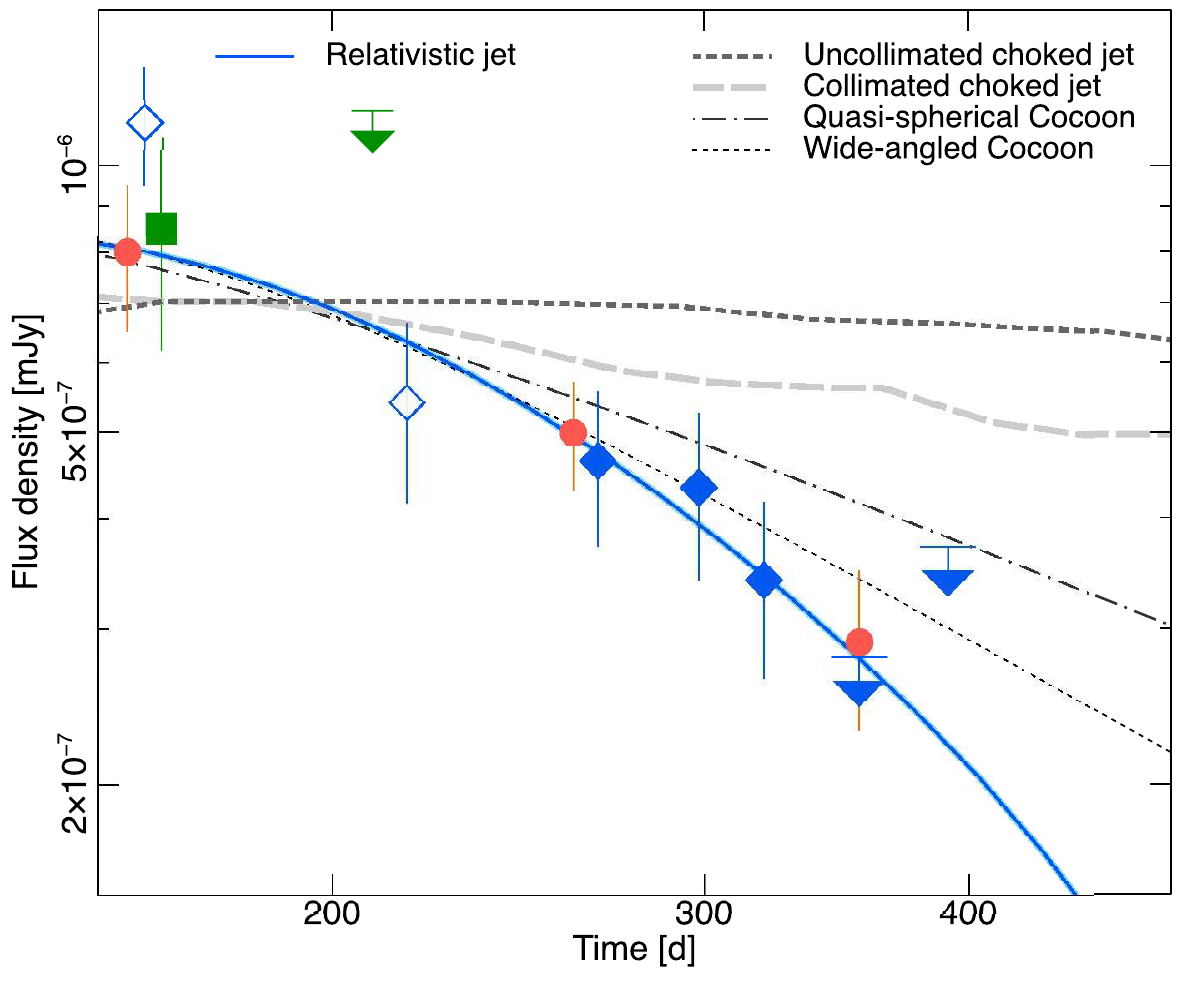}
     \caption{Late time afterglow light curves  
     compared to different explosion models: 
     choked jets from numerical simulations (thicker lines, \citealt{Nakar2018}),
     wide-angled cocoon \citep{LambMandelResmi2018}, and our best fit models
     of quasi-spherical cocoon (dot-dashed line) and structured jet (solid line). 
     Different symbols represent different wavelengths: X-rays (circles), 
     optical (downward triangles; 3\,$\sigma$ upper limits), 
     and radio (diamonds) from ATCA (filled) and VLA (empty). 
     For plotting purposes, data were rescaled to a common energy 
     of 5 keV using the observed spectral slope $\beta$=0.585. 
     }
     \label{fig:res}
     \vspace{-0.3cm}
\end{figure*}

\begin{figure*}
\includegraphics[width=0.8\columnwidth]{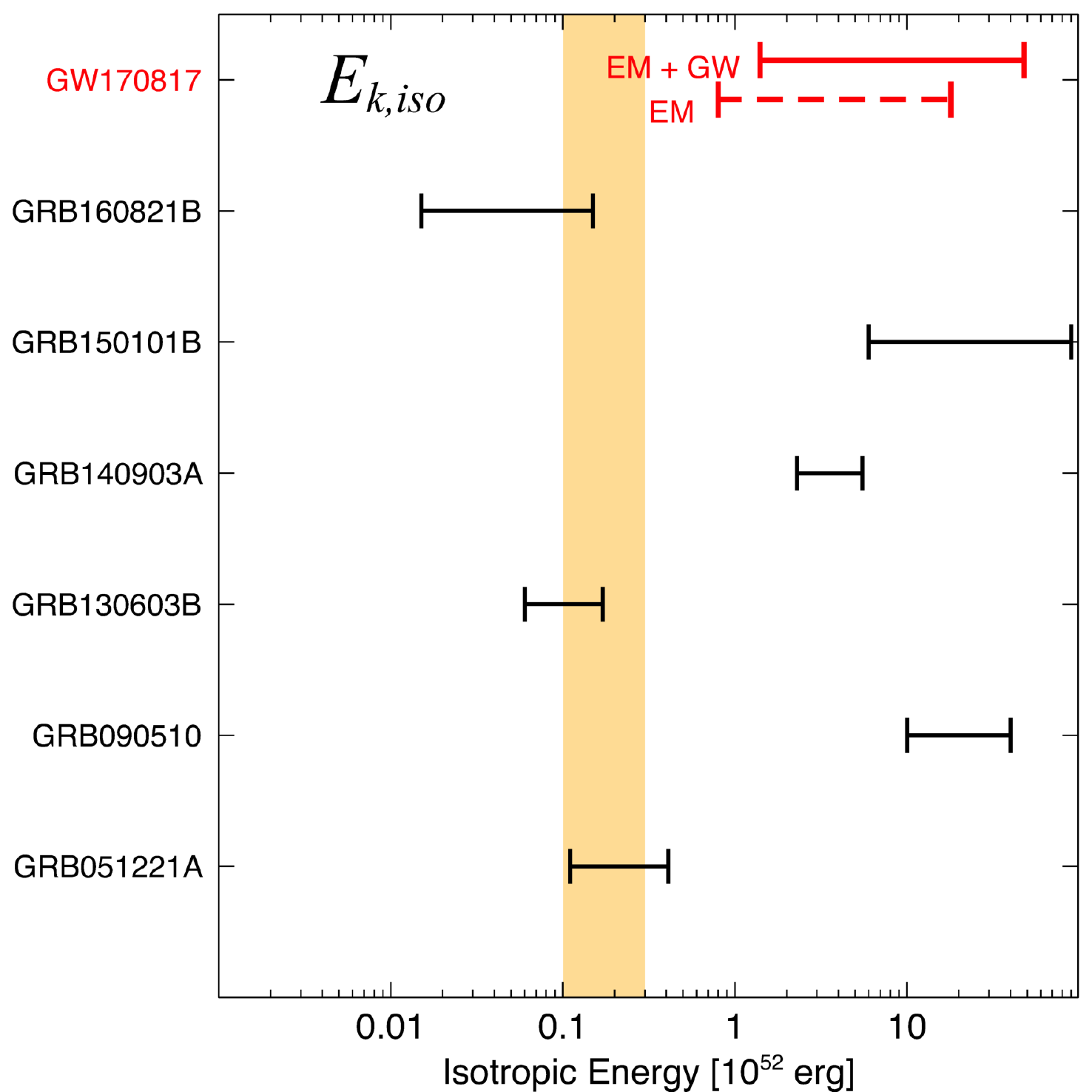}\hspace{0.6cm}
\includegraphics[width=0.8\columnwidth]{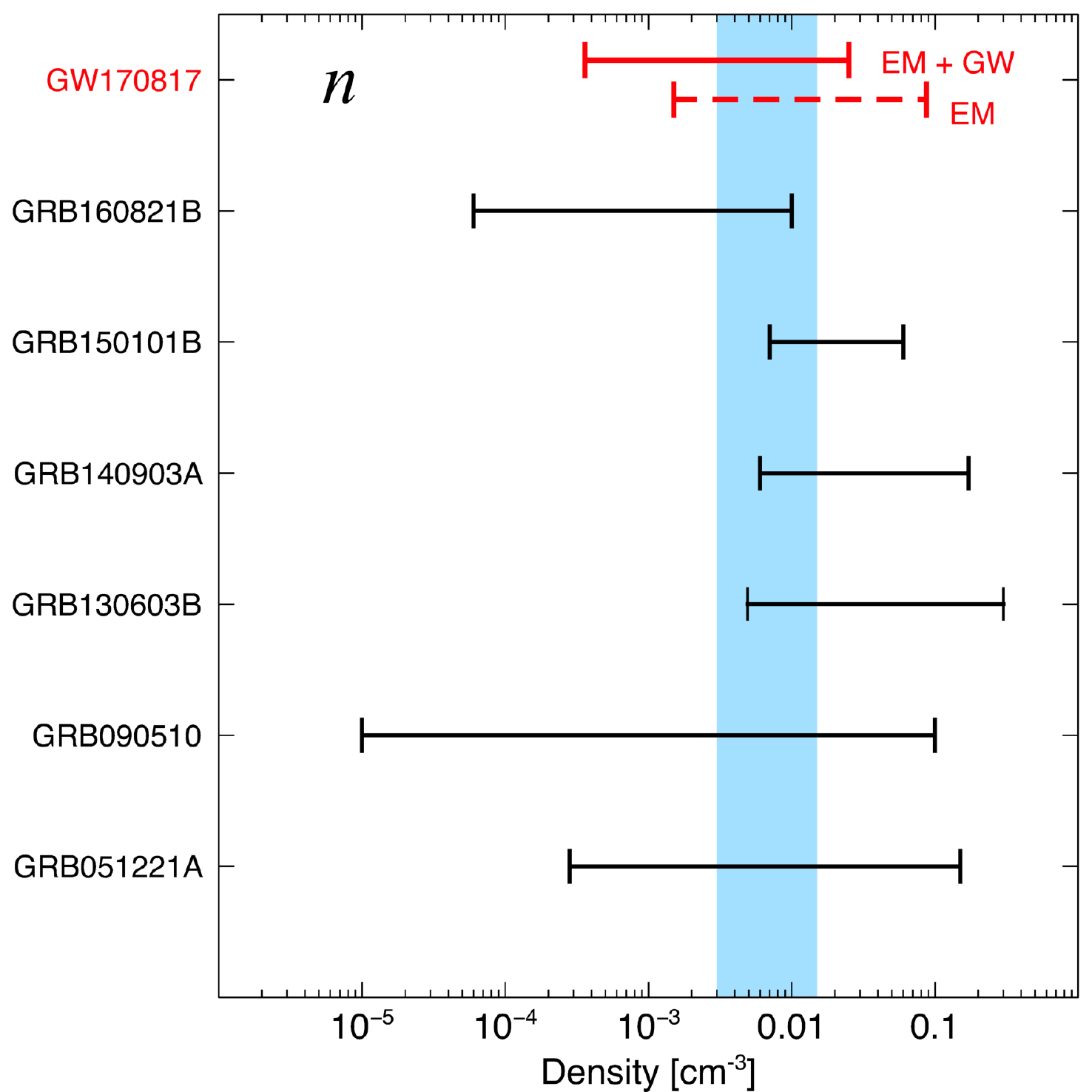}\\
\vspace{0.5cm}
\includegraphics[width=0.8\columnwidth]{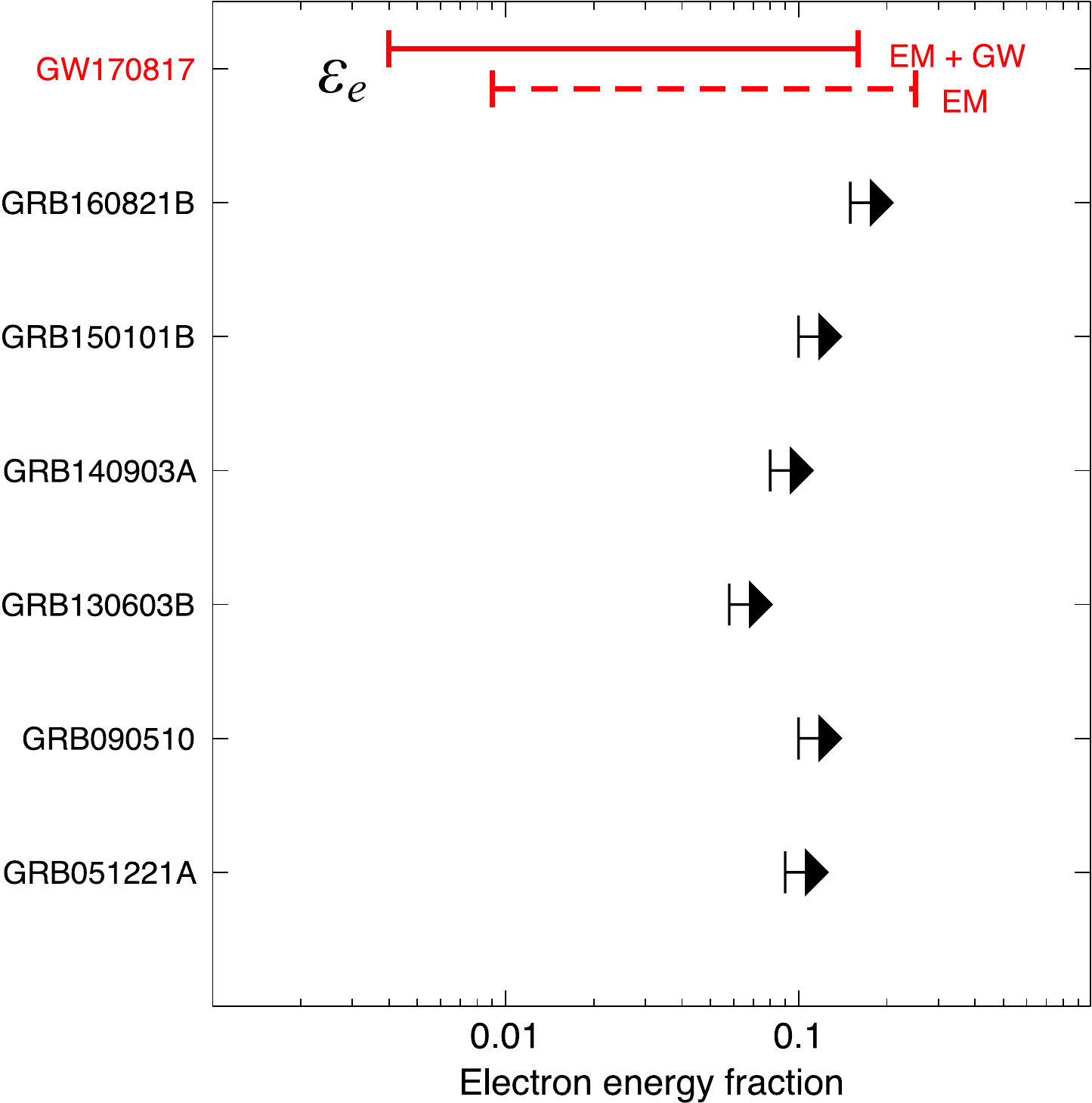}\hspace{0.6cm}
\includegraphics[width=0.8\columnwidth]{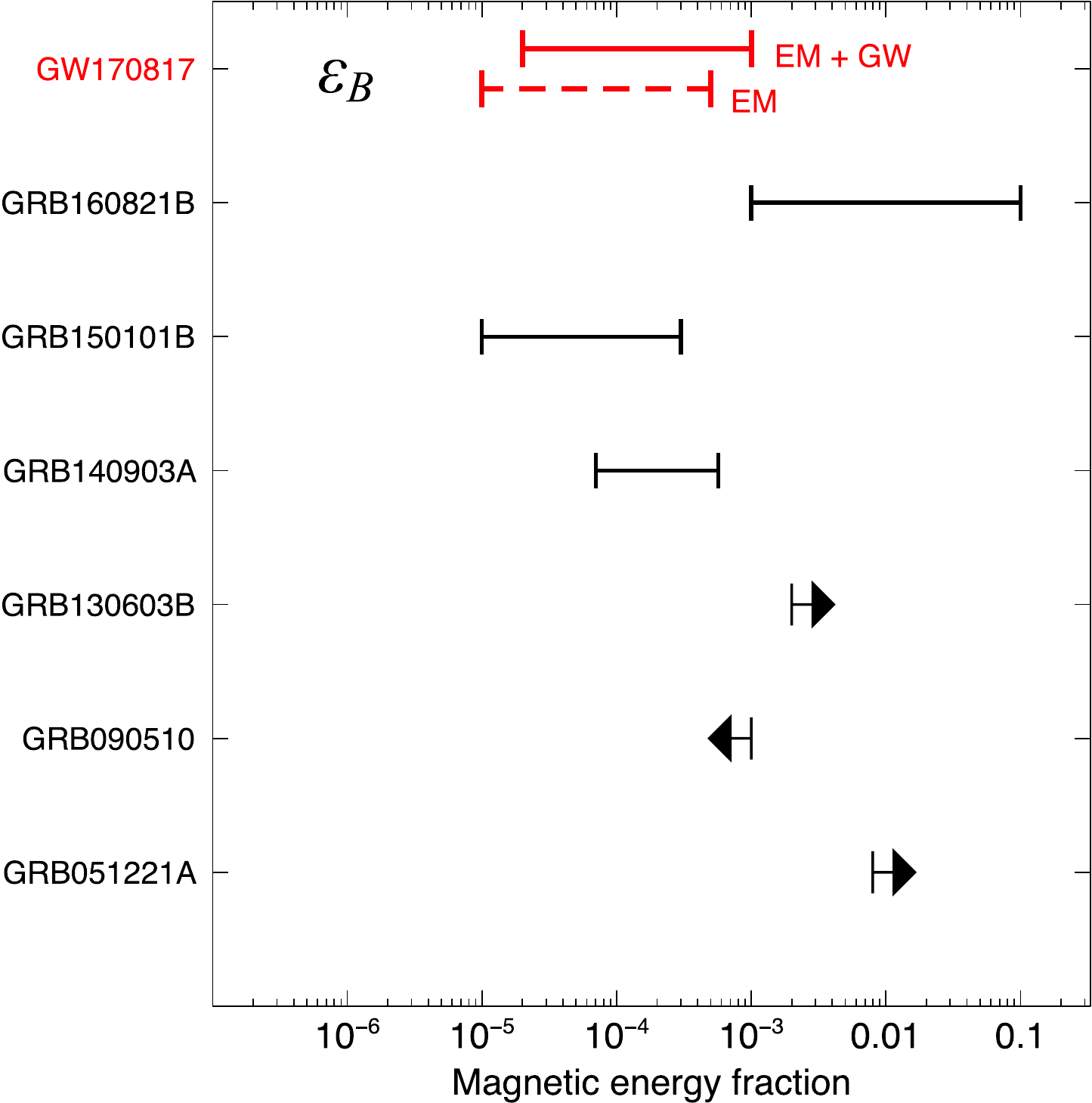}\\
\vspace{0.5cm}
\includegraphics[width=0.8\columnwidth]{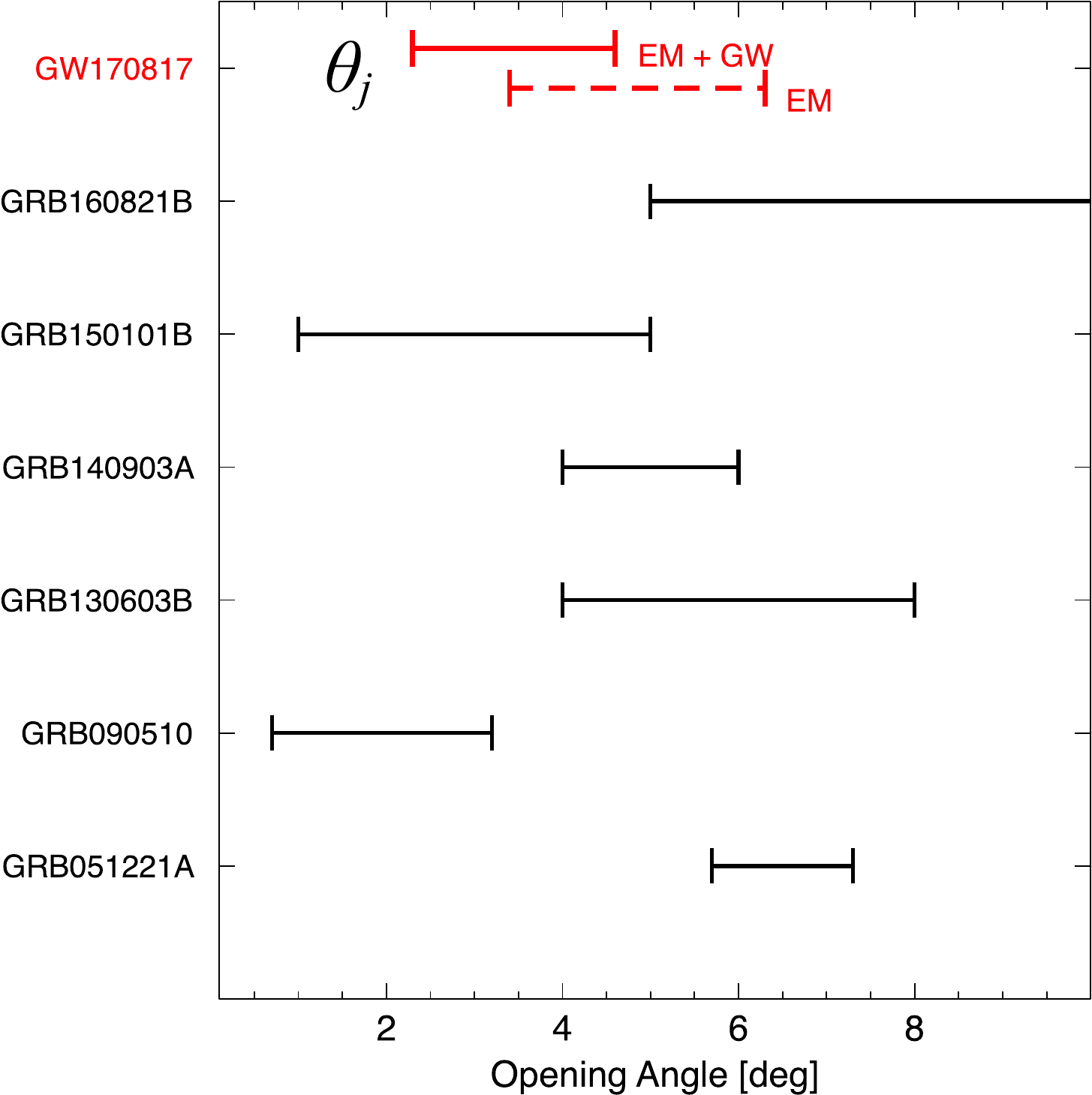}
     \caption{Afterglow parameters for GW170817 and a sample of cosmological ($z>$0.1) short GRBs
     with multi-wavelength afterglows. For GW170817 we report the results based purely on the electromagnetic 
     observations (EM), and those incorporating the LIGO constraints on the binary inclination (EM+GW). 
     Data are from: \citet{Soderberg2015,Kumar2010,Fong2015,Troja2016,KN150101B,KN160821B}
     The vertical grey band show the median values of energy $E_{k,\rm iso}$ and density $n$
     for the larger sample of {\it Swift} short GRBs \citep{Fong2015}.}
     \label{fig:sGRBs}
     \vspace{-0.3cm}
\end{figure*}

\section{Results}

\subsection{A rapid afterglow decline: constraints on the outflow structure}
\label{decline_results_section}

For jets that fail to break out (``choked jets''), the jet energy is dissipated into a surrounding cocoon of material. This scenario is therefore included in our group of `cocoon' models \citep{Troja2018}. 
The post-peak temporal slope is a shallow decay of $\alpha \approx 1.0-1.2$ up to at least 300 days for GRB 170817A, as can be inferred from semi-analytical modeling of the evolution of a trans-relativistic shell \citep{Troja2018}. Any remaining post-turnover impact due to continued energy injection from e.g. a complex velocity profile of the ejecta, would lead to an even shallower decay. 
Afterwards, the slope will eventually become that of an expanding non-relativistic (quasi-)spherical shell. 
As for the Sedov-Taylor solution of a point explosion in a homogeneous medium, this slope translates to $\alpha = (15p - 21)/10$ for $\nu_m < \nu_{obs} < \nu_c$, when combined with a standard synchrotron model for shock-accelerated electrons \citep{FrailWaxmanKulkarni2000}. The value $p \approx 2.17$ implies $\alpha \approx 1.155$. If $\nu_m, \nu_c < \nu_{obs}$, $\alpha = (3p-4)/2 \approx 1.255$ instead \citep[e. g.][]{GranotSari2002}. 
If the cocoon choking the jet is not quasi-spherical, but merely wide-angled, some sideways spreading of the outflow may still occur. By definition for a non-relativistic flow velocity, this will not produce any observational features related to relativistic beaming, but the continuous increase in working surface will give rise to additional deceleration of the blast wave relative to the case of purely radial flow. As a result, the temporal slope could steepen slightly by another $\Delta \alpha \approx 0.15 - 0.2$ \citep{LambMandelResmi2018}, before settling into the late quasi-spherical stage. For GW170817, this implies a maximum $\alpha \approx 1.35$ from a cocoon-dominated / choked jet model. \\
\indent By contrast, if the jet has a relativistically moving inner region (in terms of angular distribution of Lorentz factor), the post-peak temporal slope will be like that of an on-axis jet seen after the jet break: the entire surface of the jet has come into view and there is no longer a contribution to the light curve slope from a growing visible patch. 
Different calculations predict various degrees of steepening \citep[e.g.][]{GillGranot2018}: a slope $\alpha \approx p$ according to analytical models \citep{SariPiranHalpern1999}, and a somewhat steeper $\alpha \approx 2.5$ according to semi-analytical models \citep{Troja2018} and hydrodynamical simulations of jets (\citealt{vanEertenMacFadyen2013}, although the latter were done for jets starting from top-hat initial conditions).

As derived in section \ref{temporal_properties_section}, the decay slope $\alpha_2$ of the empirical model exceeds the predictions of most choked jet models  by a healthy margin. This is confirmed by the comparison of the late afterglow data, from radio to X-rays, with physical models
of choked and structured jets (Figure~\ref{fig:res}). 
The observed decay is consistent with the turnover of a structured jet (solid line).  While the observed $\alpha_2$ is not as steep as $2.5$, this is not unexpected for a jetted flow as the transition of the light curve from rise to decay is spread out over time (and fully captured by the direct application of a structured jet model in section \ref{modeling_section}, see also Fig.~\ref{fig:lc}). 

The rapid decline of the afterglow therefore poses an additional challenge to the choked jet models, in support to the results of the high-resolution radio imaging \citep{Mooley2018superluminal,Ghirlanda2018}. 
While growing evidence indicates that the merger remnant launched a successful relativistic jet, the presence of a cocoon cannot be excluded.  There might well be an observable cocoon component present in the outflow even for successful jets \citep{Nagakura2014, Murguia-Berthier2014}.
Indeed, the structured jet itself might be an indication of the presence of a cocoon (in case the structure is not imposed by the torus upon launching, \citealt{AloyJankaMueller2005}).
However, it would appear any such cocoon is not the dominant emission component at late times.

\subsection{Afterglow properties: comparison to short GRBs}
\label{comparison_section}

The predictive power of each model can be judged by the deviance information criteria (DIC), where lower scores correspond to greater predictive power \citep{DIC02}.  The Gaussian jet model fit has a DIC of 103.5 and the cocoon model fit has a DIC of 151.8, favoring the Gaussian jet as the more predictive model. For all models and priors, the posterior value of the electron power law slope lies around $p = 2.170 \pm 0.010$, fully consistent with the value obtained by the spectral analysis (Section \ref{spectral_properties_section}).

The cocoon model requires a small amount of relativistic ejecta with a substantial Lorentz factor $\Gamma_{max} \in [6.1,200]$ (all ranges with 68\% percent confidence) followed by an energetic tail of slower ejecta with minimum Lorentz factor $\Gamma_{min} \in [1.9, 7.48]$.  The high Lorentz factors are in tension with a choked-jet scenario, where the ejecta achieve only Newtonian velocity.
The total energy, assuming a spherical blast wave, is rather high, between $10^{51}$ and $4\times 10^{53}$~erg.  
The circumburst density and shock micro-physical parameters are very poorly constrained in this model.

The Gaussian jet (Figure~\ref{fig:lc}) has a well constrained width $\theta_c$\,=\,0.06$\pm$0.02 rad (3.4$^\circ$\,$\pm$\,1.1$^\circ$), and a total energy between $5\times 10^{49}$ and $1.4\times 10^{51}$ erg. 
The wide truncation angle is largely unconstrained.
The ambient density $n$ is constrained to be between $3\times 10^{-4}$ and $2.4\times 10^{-2}$ cm$^{-3}$. The micro-physical parameters $\epsilon_e \in [4\times10^{-3}, 0.17]$ and $\epsilon_B \in [2.7\times 10^{-5}, 10^{-3}]$ are only somewhat constrained by the model fit.
The constraint $\nu_c >$ 1 keV from the lack of spectral steepening drives the relatively small value for $\epsilon_B$, as $\nu_c \propto \epsilon_B^{-3/2}$.
 These values were derived by assuming a jet with a Gaussian angular profile, yet they are in good agreement with other estimates based on different angular structures \citep[e.g.][]{Ghirlanda2018}. 

 The viewing angle derived from the electromagnetic observations alone is
 0.52$\pm$0.16 rad (30$^\circ$ $\pm$ 9$^\circ$), consistent with the constraints
 from prompt emission \citep[e.g][]{LVCGBM, Begue:2017hg}, 
 optical and radio imaging
 \citep{Mooley2018superluminal,Ghirlanda2018, Lamb2019}.
 By adding the GW constraints on the binary inclination $\iota$ to our modeling, 
 we obtain $\theta_v$ = 0.38$\pm$0.11 rad (22$^\circ$\,$\pm$\,6$^\circ$),
 consistent with the LIGO estimates that informed the prior.
 The good agreement of the electromagnetic and GW constraints suggest that the 
 relativistic jet was launched nearly perpendicularly to the orbital plane.

The year-long monitoring of GW170817 significantly reduced the allowed parameter space of models, tightening the constraints on the afterglow properties. 
This allows for a comparison with other well-studied short GRB explosions, as presented in Figure~\ref{fig:sGRBs}. 
It is remarkable how the properties of GW170817 fit within the range of short GRB afterglows.  The low circumburst densities $n\approx$0.01\,cm$^{-3}$ are typical of the interstellar medium, and consistent with the location of these bursts within their galaxy light. The electron energy fraction seems well constrained to $\epsilon_e \gtrsim$0.1, whereas $\epsilon_B$ tends to lower values $\lesssim$0.01, and is only loosely constrained. 
The narrow width of the Gaussian jet $\theta_c$ of GW170817 is comparable to the half-opening angle $\theta_j$ inferred from top-hat jet models of short GRBs \citep[e.g.][]{Troja2016}, suggesting that these GRB jets had narrow cores of similar size.
The isotropic-equivalent energy is also consistent with the measurements from other short GRBs, although we note that most events in the sample lie above the median value of $E_{k,iso}$ (vertical band). This is not surprising as we selected the cases of well-sampled light curves with good afterglow constraints, 
thus creating a bias toward the brightest explosions. 

\section{Conclusions}

The long-term afterglow monitoring of GW170817 supports the earlier suggestions
of a relativistic jet emerging from the merger remnant, and challenges the alternative scenarios of a choked jet. 
Whereas emission at early times ($<$160 d) came from the slower and less energetic lateral wings, the rapid post-peak decline suggests that emission from the narrow jet core has finally entered our line of sight. The overall properties of the explosion, as derived from the afterglow modeling, are consistent with the range of properties observed in short GRBs
at cosmological distances, and suggest that we detected its electromagnetic emission thanks to a combination of moderate off-axis angle ($\theta_v$-$\theta_c$\,$\approx$20$^{\circ}$ ) and intrinsic energy of the explosion.

\bibliographystyle{aa}

\begin{table*}
        \centering
        \caption{Late-time X-ray and radio observations of GW170817}
        \label{tab:obs}
        \begin{tabular}{cccccccc}
        \hline
        T-T$_0$  & Facility & Exposure & $\beta$  & Flux\footnotemark & Frequency  \\
         (d) & & & & ($\mu$Jy) & (GHz) \\
        \hline
       267  &    ATCA      &   11.0h  &  0.8$\pm$0.8  &  30$\pm$7               &    5.5  \\
             &              &          &        &  20$\pm$6               &    9.0  \\
             &              &          &        &  30$\pm$6               &    7.25 \\
       298  &    ATCA      &   11.0h  &  -0.3$\pm$1.0  &  25$\pm$7               &    5.5 \\
             &              &          &        &  29$\pm$6               &    9.0  \\
             &              &          &        &  28$\pm$6               &    7.25 \\
       320  &    ATCA      &  11.5h   &   0.4$\pm$0.8  &  27$\pm$8               &    5.5  \\
             &              &          &        &  22$\pm$6               &    9.0  \\
             &              &          &        &  22$\pm$5               &    7.25 \\
       359  &    ATCA      &   9.5h   &  $--$  &  $<$26                  &    5.5  \\
             &              &          &        &  $<$20                  &    9.0  \\
             &              &          &        &  $<$18                  &    7.25 \\
       391 & ATCA &  9.5h    &  $--$  &  $<$33                  &    5.5  \\
             &              &          &        &  $<$27                  &    9.0  \\
             &              &          &        &  $<$24                  &    7.25 \\  
   \hline
      359   &  {\it Chandra} & 67.2 ks &  0.8$\pm$0.4  & 2.8$^{+0.3}_{-0.5}$ $\times$ 10$^{-4}$ & 1.2$\times$10$^9$ \\ 
             &   &  &  0.585  & 2.1$^{+0.2}_{-0.3}$$\times$ 10$^{-4}$ & 1.2$\times$10$^9$ \\ 
      \hline
    \end{tabular}
\end{table*}
    
\footnotetext{X-ray fluxes are corrected for Galactic absorption along the sightline. }

\newpage
\begin{table*}
        \centering
        \caption{Constraints on the Gaussian jet and Cocoon model parameters.  Reported are the median values of each parameter's posterior distribution with symmetric 68\% uncertainties (ie. the 16\% and 84\% quantiles).  The first Gaussian jet column uses constraints from the afterglow alone, the second includes the LIGO constraints on the inclination angle using the Planck value of $H_0$.}
        \label{tab:MCMC}
    \begin{tabular}{cc}
        \begin{tabular}{lcc}
                \hline
        Parameter & Jet & Jet+GW+Planck \\
        \hline
        $\thetaobs$ &           $0.52^{+0.16}_{-0.16}$  
                    &           $0.38^{+0.11}_{-0.11}$ \\[3pt]
        $\log_{10}E_0$ &        $52.47^{+0.81}_{-0.56}$   
                        &       $52.80^{+0.89}_{-0.65}$ \\[3pt]
        $\theta_c$ &            $0.079^{+0.026}_{-0.024}$ 
                        &       $0.059^{+0.017}_{-0.017}$ \\[3pt]
        $\theta_w$ &            $0.77^{+0.47}_{-0.38}$  
                        &       $0.61^{+0.42}_{-0.31}$ \\[3pt]
                        \\
        \hline
        $\log_{10} n$ &       $-1.83^{+0.77}_{-1.0}$  
                        &       $-2.51^{+0.90}_{-0.99}$\\[3pt]
        $p$ &                   $2.1678^{+0.0064}_{-0.010}$ 
                        &       $2.1681^{+0.0062}_{-0.0095}$ \\[3pt]
        $\log_{10}\epse$ &      $-1.13^{+0.53}_{-0.88}$  
                        &       $-1.39^{+0.62}_{-0.99}$ \\[3pt]
        $\log_{10}\epsB$ &      $-4.18^{+0.85}_{-0.58}$  
                        &       $-4.00^{+1.0}_{-0.69}$ \\
        \hline
        $\log_{10} E_{tot}$ &   $50.24^{+0.72}_{-0.47}$  
                        &       $50.30^{+0.84}_{-0.57}$ \\
                \hline
        \end{tabular} & \begin{tabular}{lc}
        \hline
        Parameter & Cocoon \\
        \hline
        $\log_{10} \umax$ &     $1.38^{+0.92}_{-0.6}$ \\[3pt]
        $\log_{10} \umin$ &     $0.51^{+0.36}_{-0.29}$  \\[3pt]
        $\log_{10} \Einj$ &     $56.1^{+3.2}_{-2.9}$\\[3pt]
        $k$               &     $7.26^{+0.41}_{-0.57}$ \\[3pt]
        $\log_{10} \Mej$ &      $-8.2^{2.0}_{-1.3}$ \\
        \hline
        $\log_{10} n$ &       $-4.5^{+2.1}_{-2.4}$ \\[3pt]
        $p$ &                   $2.1715^{+0.0055}_{-0.0057}$ \\[3pt]
        $\log_{10}\epse$ &      $-2.0^{+1.3}_{-1.3}$ \\[3pt]
        $\log_{10}\epsB$ &      $-3.4^{+1.7}_{-1.1}$ \\
        \hline
        $\log_{10} E_{tot}$ & $52.4^{+1.2}_{-1.3}$ \\
        \hline
        \end{tabular}
    \end{tabular}
\end{table*}

\begin{figure*}
\includegraphics[width=1.5\columnwidth]{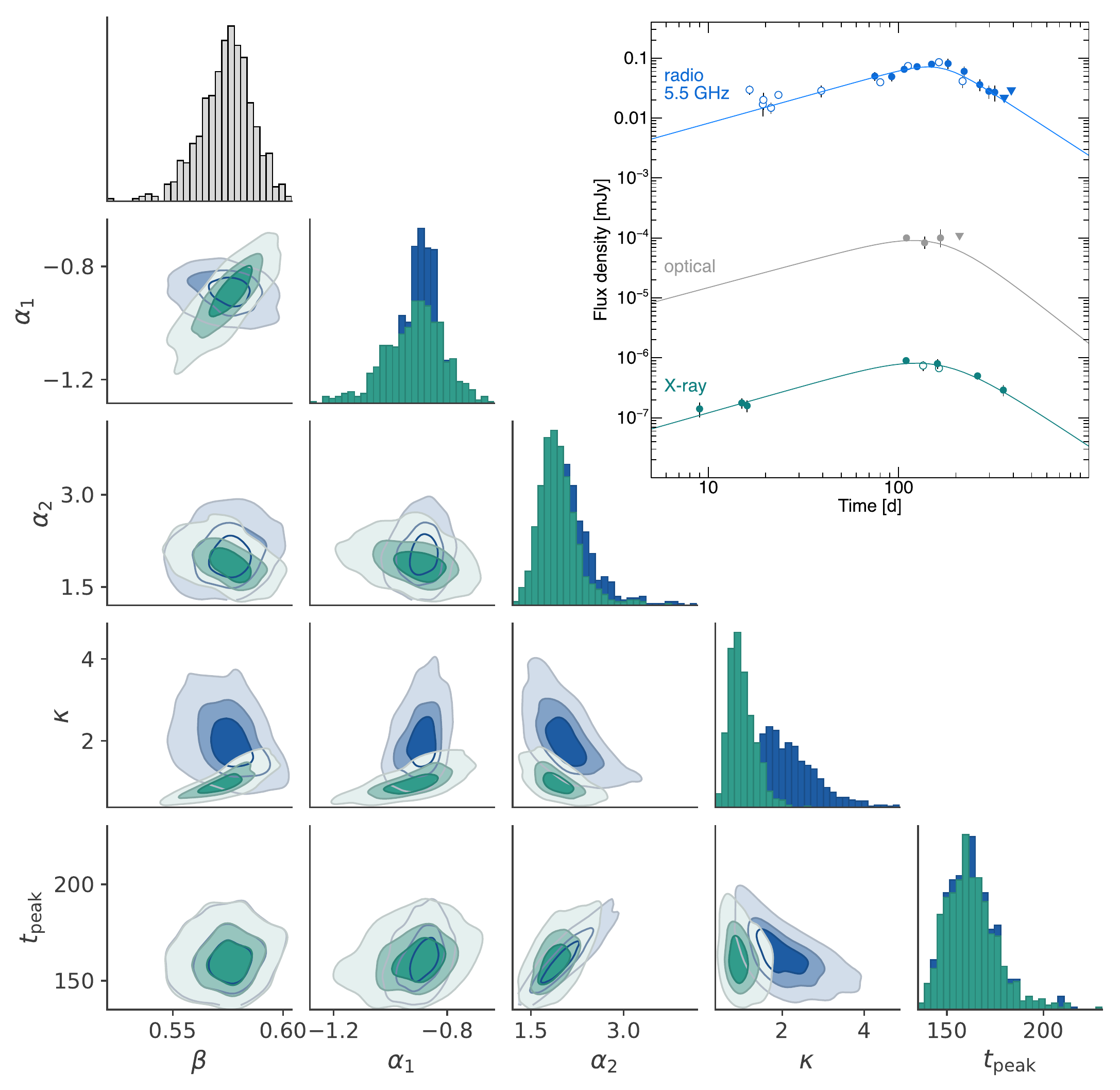}
     \caption{Corner plot for the empirical model described in
     Sect. 2.2.  
     { The prior on each of these parameters is normal, the center
     and standard deviation were: $\beta$ = (1,1), $\alpha_1$ = (0.5,1), 
     $\alpha_2$ = (-2.5,1), log $\kappa$ = (0,0.5),  log $t_{\rm peak}$ = (2,2).}
     We report the confidence contours derived from radio { (blue)} and X-ray { (green)} data, as the optical light curve is only poorly constrained. 
     The best fit model is shown in the top-left corner.
     }
     \label{fig:sup1}
     \vspace{-0.3cm}
 \end{figure*}

\begin{figure*}
\includegraphics[width=1.95\columnwidth]{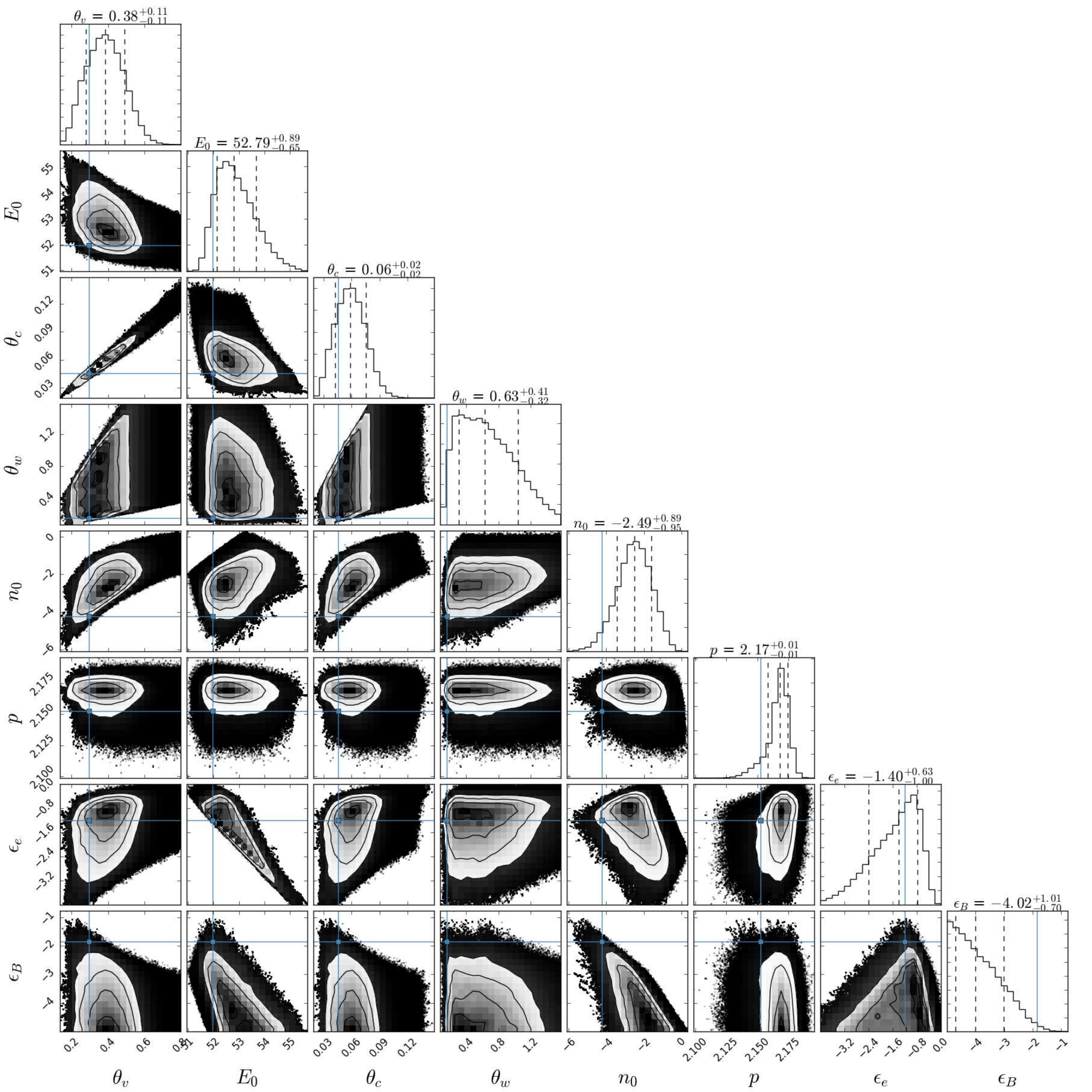}
     \caption{ Fit result for the Gaussian jet model. This ``corner plot'' shows all one-dimensional (diagonal) and two-dimensional (off-diagonal) projections of the posterior probability density function. The best-fit value (maximum posterior probability) is shown in blue. Dotted lines mark the 16\%, 50\%, and 84\% quantiles of the marginalized posteriors for each parameter.}
     \label{fig:corner}
\end{figure*}

%%%%%%%%%%%%%%%%%%%%%%%%%%%%%%%%%%%%%%%%%%%%%%%%%%%%%%%
\bsp
\label{lastpage}
\end{document}